\documentclass[12pt]{article}

\usepackage[russian]{babel}               % Русификация
\usepackage{graphicx,graphics}
\begin{document}
\begin{center}
{\large\bf Vector Theory of Gravity: quantum and classical effects, renormalization} \vskip 0.3 true in
{\large V. N. Borodikhin} \vskip 0.3 true in {\it
Omsk State University, pr. Mira 55a, Omsk, Russia.}
\end{center}
%\date{\today}
\begin{abstract}%
In this work, we make quantization of gravitation interaction within the framework of a vector
 theory of gravitation for the first time. The work demonstrates that this theory meets
  the requirement of renormalizability.
Here we consider some quantum effects, particularly graviton scattering on fermion and corrections
  to the Newton's. Gravitational energy sign changing mechanism and classical transition on small scale is discussed.
 It is shown that within this theory black holes of Schwarzschield hole type can exist.
  Problem of dark energy structure and acceleration of Universe expansion is investigated.
  We also consider the behavior of binary pulsars in the vector theory of gravitation.
\end{abstract}
\vskip 0.2 true in e-mail: borodikhin@inbox.ru

\section{Introduction}

In \cite{1} we make an attempt to describe
the gravitational phenomena using the vector field
approximation in Minkowski space.
Some attempts to describe the
gravity using  vector models were made earlier
 \cite{0,00}, however a number of difficulties arise in this approach,
 associated with the divergence of the previously calculated in the framework of this model
 values of the main gravitational experiments and observation.
It should be noted however, that all the calculations mentioned effects within a framework of the
 vector gravity theory performed either by neglecting the amendments  related to the vector field, or in
  Newtonian approximation.

In \cite{1} we have calculated the values of the Mercury orbit
perihelion shift, the light deflection angle in the gravitational
field of the Sun and the radar echo delay in
a post-Newtonian approximation. The values found
coincide with the experimental ones.
It is shown
that, in this theory, there exists a model of an expanding Universe.

In this work, we make quantization of a vector gravitation field.
 It is demonstrated that the gravitons have spin 1 and positive energy for
 the first time.

In this work, we construct the quantum theory of vector gravitation for fermions and gravitons.

We examine some quantum effects, particularly graviton scattering on fermion and corrections to the Newton's
 law. We demonstrate that on quantum level not only
  the gravitation attraction force, but also the gravitation repulsive force occurs in the first order of
   the perturbation theory. We also demonstrate that the gravitation constant grows with the distance decrease.

It is shown that such a theory is renormalizable.

Gravitational energy sign changing mechanism and traditional transition on small scale is discussed.

Also in the article deals with the problem of dipole gravitational radiation of binary pulsar
 and peculiarities of their structure. The mechanisms of compensating dipole
 gravitational radiation are discussed. A new mechanism of compensation on the basis of
 inhomogeneous cyclic field is proposed.

It is shown that within this theory black holes of Schwarzschield hole type can exist.
 It is also shown that it is possible to identify gas gravitino with negative pressure
  and dark energy caused by the appreciable acceleration of Universe expansion.

\section{Quantization of gravitation}

Let us consider the general theory of the vector gravitational field that is based on the Minkowski space\cite{1}.
We will connect the gravitational field with the 4-potential $A^i=(\varphi, c\vec A)$, where
$\varphi$- is the usual scalar potential and $\vec A$ is a vector potential, and $c$ is the speed of light.
The Lagrangian of the gravitational field with
account for matter has the form

\begin{equation}\label{5}
\L=-A_ij^i+\frac{1}{16\pi\gamma}G_{ik}G^{ik},
\end{equation}

where $\gamma$ is the gravitation constant,
$j^i=\mu\frac{1}{c}\frac{dx^i}{dt}$ is
the mass current density vector
, $\mu$- is the mass density
of bodies, and
$G_{ik}=\frac{\partial A_k}{\partial x^i}-
\frac{\partial A_i}{\partial x^k} $ is the antisymmetric tensor of the gravitation field.

The first term describes interaction of the field and
matter, the second one characterizes the field without
particles.

The classical theory of vector gravity correctly describes all the major
 gravitational effects is constructed in \cite {1}.
This theory satisfies equivalence principle within the accuracy of the available experiments \cite{000}.

  We carry out the quantization of the gravitational field.

Due to the sign in addend in  (\ref{5}) the gravitational waves must have negative energy.

In \cite{2}, by analogy with the concept of Dirac sea, which is built on the basis of hole theory of fermions
developed the concept of boson sea implying the existence of bosons - "holes".
According to this theory the boson vacuum is filled with bosonic states with negative energy.
At the same time as the hole or fermionic antiparticles have positive energy
 bosonic holes also have positive energy.
To obtain a positive definite scalar product and construction of the Hilbert
 space for such a theory in \cite{2} used a non-local method. In \ cite {3} to obtain a positive
 definite scalar product used $\varepsilon$ - regularization.

Based on the above mentioned concept of boson sea feasible quantization
 of the vector gravitational field.
We assume that gravitons is truly neutral.

Let us write the initial Lagrangian as:

\begin{equation}\label{40}
  \L=\frac{1}{8}
  \frac{\partial A_m}{\partial x^n}\frac{\partial A^m}{\partial x_n},
\end{equation}

In (\ref{40}) introduced redesignation $A_i=A_i^{'}/\sqrt{\pi G}$ - is the vector gravitational
 field for convenience divided by
$\sqrt{\pi \Gamma}$, $\Gamma$ - is the gravitation constant.

The components of vector potential are the independent values. The potential can
 be split into positive and negative frequency compositions
$A_n(x)=A_{n}^{+}(x)+A_{n}^{-}(x)$. Further, let us pass to the impulse representation:

\begin{equation}\label{41}
A_{n}^{\pm}(x)=\frac{1}{(2\pi)^{3/2}}\int\frac{d\vec k}{\sqrt{2k_0}}e^{\pm ikx}
A_{n}^{\pm}(\vec k)
\end{equation}

For diagonalization in impulse representation we make an expansion in components
 connected with a local frame.

\begin{equation}
A_{n}^{\pm}(x)=e_{n}^{1}a_{1}^{\pm}(\vec k)+e_{n}^{2}a_{2}^{\pm}(\vec k)+
e_{n}^{3}a_{3}^{\pm}(\vec k)+e_{n}^{0}a_{0}^{\pm}(\vec k)
\end{equation}

$\vec e_1(\vec k)$ and $\vec e_2(\vec k)$ - are spatial transverse unitary vector, $\vec
e_3=\vec{k}/\vert\vec{k}\vert$ - is a longitudinal unitary vector,where

\begin{equation}
e_0^i=0, \ \  \  \vec e^i\vec e^j=\delta_{ij}, \  \  \ [\vec e^i\times\vec e^j]=\vec e^k
\  \  (i,j,k=1,2,3),
\end{equation}

$e^0$ - is a unitary time four-vector: $e_n^0=\delta_{0n}$.

Following the general scheme of quantization of massless vector field \cite{4}
 we receive commutation relations for the gravitational field:

\begin{equation}\label{43}
[a_m^{-}(\vec k)a_n^{+}(\vec k^{\prime})]_{-}=g^{mn}\delta(\vec k-\vec k^{\prime})
\end{equation}

where $g^{mn}$ - is a metric tensor.

Quantization (\ref{43}) enables us to consider the operatorors $a_n^{\pm}$ an as the creation and
 annihilation operators of different types of gravitons: time, transverse and longitudinal ones.

However, for transverse and longitudinal components in the right side there is a minus sign.
To overcome this difficulty,
 introduce an indefinite metric according to \cite{2}.
   Commutation relations become usual:

\begin{equation}\label{45}
[a_m^{-}(\vec k)a_n^{+}(\vec k^{\prime})]_{-}=\delta(\vec k-\vec k^{\prime})\delta_{nm}
\end{equation}

Under quantization the operators  $A_n$ are assumed to be independent,
 therefore, Lorentz conditions cannot be stipulated on  $A_n$.

Passing to impulse representation, let us formulate the Lorentz condition
 for acceptable states in a "slack" form:

\begin{eqnarray}\label{47}
(a_0^{-}-a_3^{-})\Phi=0
\end{eqnarray}

and conjugate:

\begin{eqnarray}\label{48}
\Phi^{\dagger}(a_0^{+}-a_3^{+})=0
\end{eqnarray}

These conditions implement the Lorentz condition on the average,
 which suffices to accordance with a classical field.

\begin{equation}\label{49}
\langle a_3^{+}a_3^{-}-a_0^{+}a_0^{-}\rangle=0
\end{equation}

 Let us write the energy-momentum vector:

\begin{equation}\label{50}
\langle \vec P^n\rangle=\int d\vec kk^n\langle a^{+ m}(\vec k)a_m^{-}(\vec k)\rangle= -\int
d\vec kk^n\langle a_1^{+}(\vec k)a_1^{-}(\vec k)+a_2^{+}(\vec k)a_2^{-}(\vec k)\rangle.
\end{equation}

Thus the energy density of gravitons is determined by the contribution of only the transverse components in accordance with the theory of boson sea, has an average
 a positive value.

In view of the average Lorentz condition, we find that the graviton spin is
 equal to $s=1$, scalar and longitudinal components are excluded.

It is commonly supposed that spin of graviton must be equal to two. It is connected with the fact that
 common traditional scalar and vector gravitation theories do not give correct results for values
  of the main gravity experiments such as deflection of light in solar field or Mercury’s perihelion
   calculation. However as shown in \cite{1} if we take lagrangian of vector
  gravito-cyclical field in the form of relativistic Esseen's lagrangian
  we can get the values of the main gravity experiments coincident
        with the observations according to the vector gravitation theory.
         All in all nothing inhibits existence of spin gravitons 1 \cite{26}.
          Using of modern units such as LISA or VIRGO has not made it possible jet to detect directly
  gravitational waves yet, and consequently define spin graviton. After all the question of spin
   graviton should be solved on the base of direct observations.

\section{Quantum Theory of Gravity}

Let us consider the density of Lagrangian describing the massless gravitons and standard fermions
 of the spin $1/2$.

  Lagrangian describing the classical vector gravitational fields is examined in \cite{1}.

\begin{equation}\label{101}
 \L= i\overline{\psi}\gamma^{\mu}\partial_{\mu}\psi-m\overline{\psi}\psi-
 g A^{\mu}\overline{\psi}\gamma^{\mu}\psi +\frac{1}{4}
  (\partial_{\mu}A_{\nu}-\partial_{\nu}A_{\mu})^2+\frac{1}{2}(\partial_{\mu}A_{\mu})^2,
\end{equation}

where $\gamma^i$ is Dirac matrixes, $\psi$ - is functions of fermion fields.

\begin{equation}\label{100}
 g=\sqrt{\pi \Gamma}m
\end{equation}

  is the gravitation charge, $m$ is the mass.

 When such a change the dimension of the field $[A_{\mu}]=\frac{d}{2}-1$,
dimension of the spinor components of the current standard: $[\psi]=\frac{d-1}{2}$.
To the third term in (\ref{101}) has the correct dimension multiply it
 on the $\xi^{2-d/2}$, where $\xi$ - arbitrary mass, d - dimension of the space

Nevertheless, it may be noted that the interaction between photons and gravitons
 as gravitons with all the moving particles exists, since the graviton
 although massless particle still has a pulse and
hence can interact with any particle having a momentum
 due to the previously mentioned vector (cyclic) interaction.

Let us write the Feynman diagrams for quantum gravitation.
The factor
$ig(\gamma^{\nu})(2\pi)^4\delta^4(p-p^{'}+k)$, where $\gamma^{\nu}$ is Dirac gamma-matrixes,
assigns to the knot with a summation index $\nu$ wherein the fermion line with impulse
 $p$ and graviton line with impulse $k$ enter and fermion line $p_{'}$ gets out.
Bispinors according to \cite{4} assign to the external fermion lines.
 $\frac{e_{\mu}^{\nu}}{(2\pi)^{3/2}\sqrt{2k_0}}$ $(\nu\ne 0)$
assigns to the graviton in initial or finite state with polarization $e_{\nu}$ and impulse $k$.
$\frac{1}{i(2\pi)^4}\int\frac{m+\hat{p}}{m^2-p^2-i\varepsilon}dp$
corresponds to the internal fermion line of particle with impulse $p$ or antiparticle  with impulse $-p$.
$\frac{ig^{\mu\nu}}{(2\pi)^4}\int\frac{dk}{k^2+i\varepsilon}$
corresponds to the internal graviton line.  The sign of a graviton propagator is opposite to the sign of
 a photon propagator.

\begin{figure}
%\hspace{-2.5cm}
 \centering
\includegraphics[width=11cm,height=1cm]{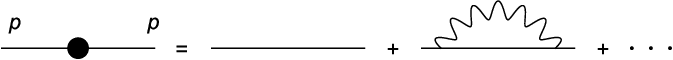}\\
\caption{Expansion of the fermion propagator.}% Подпись рисунка
%\label{pic} % Метка для ссылки на рисунок.
\end{figure}

\begin{figure}
%\hspace{-2.5cm}
 \centering
\includegraphics[width=14cm,height=3cm]{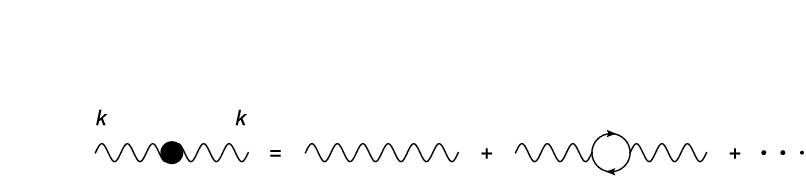}\\
\caption{Expansion of the graviton propagator. }% Подпись рисунка
%\label{pic} % Метка для ссылки на рисунок.
\end{figure}

\begin{figure}
%\hspace{-2.5cm}
 \centering
\includegraphics[width=12cm,height=3cm]{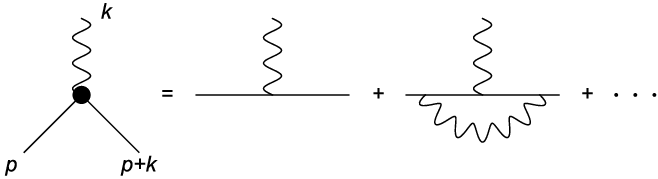}\\
\caption{ Expansion of the vertex function. }% Подпись рисунка
%\label{pic} % Метка для ссылки на рисунок.
\end{figure}

Let us determine the fermion propagator (fig.1.):

\begin{equation}\label{102}
iS_{F}^{'}=iS_{F}+iS_{F}\frac{\Sigma}{i}iS_{F}+iS_{F}\frac{\Sigma}{i}iS_{F}\frac{\Sigma}{i}iS_{F}+...=
iS_{F}(1+\frac{\Sigma}{i}iS_{F}^{'}),
\end{equation}

wherefrom

\begin{equation}\label{103}
S^{'-1}_{F}=S_{F}^{-1}-\Sigma,
\end{equation}

where $\Sigma$ corresponds to the rest energy of a fermion.
 Let us determine the graviton propagator (fig.2.):

\begin{equation}\label{104}
D^{'}(k)=D(k)+D(k)\Pi(k)D(k)+D(k)\Pi(k)D(k)\Pi(k)D(k)+...
\end{equation}

wherefrom

\begin{equation}\label{105}
D^{'}(k)^{-1}=D(k)^{-1}-\Pi(k),
\end{equation}
where $\Pi(k)$ is the rest energy of a graviton or vacuum polarization.

Let us determine the vertex function  (fig.3.):

\begin{equation}\label{106}
\Gamma_{\mu}(p,q,p+q)=\gamma_{\mu}+\Lambda_{\mu}(p,q,p+q),
\end{equation}

We have equality from the Ward identity:

\begin{equation}\label{107}
-\frac{\partial\Sigma}{\partial p^{\mu}}=\Lambda_{\mu}(p,0,p).
\end{equation}

Values $\Sigma$, $\Pi$ и $\Lambda$ are defined by the relevant Feynman diagrams.

Let us introduce the renormalized values:

\begin{equation}\label{108}
\psi_B=Z_2^{1/2}\psi, \\
A^{\mu}_B=Z_3^{1/2}A^{\mu}, \\
g_B=Z_3^{-1/2}g, \\
m_B=m-\delta_m,
\end{equation}

where index $B$ corresponds to the "naked" priming values.
 Total naked Lagrangian of quantum gravitation takes the form of:

\begin{equation}\label{109}
 \L= iZ_2\overline{\psi}\gamma^{\mu}\partial_{\mu}\psi-(m-\delta_m)\overline{\psi}\psi-
 Z_2\mu^{\varepsilon/2} g A^{\mu}\overline{\psi}\gamma^{\mu}\psi +\frac{Z_3}{4}
  (\partial_{\mu}A_{\nu}-\partial_{\nu}A_{\mu})^2 + gauge \  \ terms,
\end{equation}

where $\mu^{\varepsilon/2}$ is the dimensional parameter.

\section{Graviton scattering on fermion}

Let us consider the process of graviton scattering on fermion. Diagrams of the process
 take the following form  (fig. 4a, 4b).
The process is practically identical with the Compton effect \cite{5},
but here the gravitons are involved instead of the photons.

\begin{figure}
%\hspace{-2.5cm}
 \centering
\includegraphics[width=8cm,height=5cm]{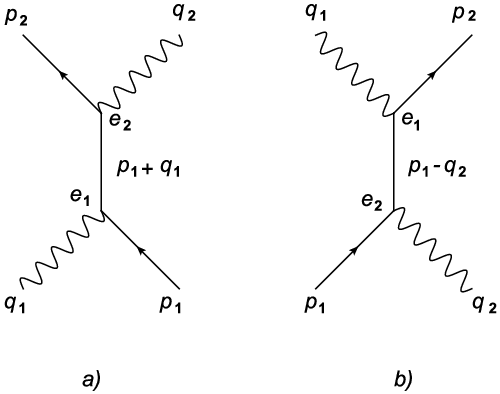}\\
\caption{ Graviton scattering on fermion. }% Подпись рисунка
%\label{pic} % Метка для ссылки на рисунок.
\end{figure}

Differential cross section of scattering can be written as:

\begin{equation}\label{188}
d\sigma=\frac{\omega_2^2}{64\pi^2m^2c^2\omega_1^2}\vert M\vert^2d\Omega,
\end{equation}

where $d\Omega$ - is the element of a solid angle, $M$ is the matrix elements of the process,
$\omega_1$, $\omega_2$ are the start and final frequencies of gravitons.

Matrix element for the first process takes the following form:

\begin{equation}\label{189}
-ig^2\widetilde{u}_2\widehat{e}_2\frac{1}{\widehat{p}_1-\widehat{q}_1-m}\widehat{e}_1u_1,
\end{equation}

where $\widetilde{u}$ and $u$ are the bispinors of particles and antiparticles.
For the second process:

\begin{equation}\label{190}
-ig^2\widetilde{u}_2\widehat{e}_1\frac{1}{\widehat{p}_1+\widehat{q}_2-m}\widehat{e}_2u_1,
\end{equation}

It is necessary to take into account both contributions.

Consequently for cross section of scattering of plain-polarized gravitons, we obtain:

\begin{equation}\label{191}
d\sigma=\frac{g^2\omega_2^2}{4c^4\omega_1^2}\Bigl[\frac{\omega_2}{\omega_1}
+\frac{\omega_1}{\omega_2}-2+4(e_1e_2)^2\Bigr]d\Omega,
\end{equation}

We can estimate the order of section. If the mass corresponds to the electron mass, section has the order
 $\sim 10^{-115}$ square meters.  If the mass corresponds to the neutron mass, without the contribution of
  strong interactions, the section has the order  $\sim 10^{-108}$.
With the Plank mass, section has the order $\sim 10^{-70}$.

\section{Quantum corrections to the Coulomb's low of gravitation}

Let us give consideration to the corrections to the Newton's low of gravitation
 in one-loop approximation connected with the corrections to the rest energy of graviton  (\ref{105}).
These corrections due to the fact that the changed sign of the gravitational energy
 are practically identical to the corrections in quantum electrodynamics to the Coulomb's
 law, and contrast with them only by the interaction constant  \cite{8}.

With provision for quantum corrections we write the scalar potential in the impulse space:

\begin{equation}\label{111}
\varphi(k)=\varphi_0(k)+\Pi(-k^2)\varphi_0(k).
\end{equation}

where $\varphi_0(k)$ is the unperturbed potential.

As a result, we obtain for a particle of mass $m_1$ with the corresponding gravitational charge $g_1$:

\begin{equation}\label{112}
\varphi(r)=\frac{g_1^2}{\pi m_1 r}\Bigl[1+\frac{2g^2}{3\pi}\int\limits_{1}^{\infty}
e^{\frac{2r\zeta}{\lambda}}\Bigl(1+\frac{1}{2\zeta^2}\Bigr)\frac{\sqrt{\zeta^2-1}}{\zeta^2}d\zeta \Bigr],
\end{equation}
where $\lambda=h/mc$.

Let us consider two limiting cases:

1. $r/\lambda\gg 1$.\\
Making substitutions  $\zeta=1+\xi$ we get:

\begin{equation}\label{113}
\varphi(r)=\frac{g_1^2}{\pi m_1 r}\Bigl[1+\frac{g^2\sqrt{\pi}}{4\pi}
\frac{e^{\frac{2r}{\lambda}}}{(r/ \lambda)^{3/2}}\Bigr].
\end{equation}

2. $r/\lambda\ll 1$.\\
\begin{equation}\label{114}
\varphi(r)=\frac{g_1^2}{\pi m_1 r}\Bigl[1+\frac{2g^2}{3\pi}\Bigl(\ln{\frac{\lambda}{r}-C-\frac{5}{6}}\Bigr)\Bigr],
\end{equation}

where $C=0.577..$ is the Euler's constant.

Therefore, polarization of the gravitational vacuum distorts
 the field of point mass in area
 $r\leq \lambda$. Outside this area perturbed repulsion field decays under the exponential law.
    In consequence of the weakness of gravitation constant and the smallness of elementary
     particles mass, this effect is nearly invisible. In area  $\sim m_{Pl}$ of the order of Plank mass,
      in very small distances this effect can play more prominent role.

     Asymptotical behavior of the  "purely gravitational charge " is:

\begin{equation}
g^2(\mu)=\frac{g^2(\mu_0)}{1-\frac{g^2(\mu_0)}{8\pi^2}\ln\frac{\mu}{\mu_0}}.
\end{equation}

\section{Renormalizability}

Let us consider the question of renormalizability vector gravitational field.

We can show that according to the
       classification \cite{4} vector gravitational interactions can be referred
        to the class of renormalizable interactions meeting the condition
         of necessary dimension of constant of maximal index of vertex
  $\omega_i\leq 0$, for the field of spin $1$.

The general formula for conditional degree of divergence  $D$ of Feynman diagrams of quantum
 theory of gravitation in $d$ spacetime dimensions has the form of :

\begin{equation}\label{121}
D=d+n\Bigl(\frac{d}{2}-2\Bigr)-\Bigl(\frac{d-1}{2}\Bigr)E_e-\Bigl(\frac{d-2}{2}\Bigr)P_e,
\end{equation}

where $P_e$ is the number of internal graviton lines,
$E_e$ is the number of internal fermion lines, $n$ is the number of vertexes.
For $d=4$ we get:

\begin{equation}\label{122}
D=4-\frac{3E_e}{2}-P_e.
\end{equation}

It is obvious that  $D$ does not depend on  $n$, which is a necessary condition for renormalizability.

Let us prove that quantum theory of gravitation is renormalizable in all orders of the perturbation theory.
 The proof is practically identical to the proof of renormalizability of quantum electrodynamics  \cite{9,10},
and reproduce it in this article there is no need.

\section{The classical limit}

Let’s consider the back transition from quantum gravitational field to classical one.
 In the limit of large quantum numbers of the oscillators, obeying Bose statistics can be neglected in the
  commutation relations of the unit, then

\begin{equation}
a_m^{-}a_n^{+}\sim a_n^{+}a_m^{-},
\end{equation}

and so these operators pass to commuting classical values defining classical densities of field.
 At that it is essential to average density of field value by a short interval of time $\Delta t$
in order to exclude infinite value of energy appearing at summing over all the energy field conditions.
 Only the frequencies meeting the condition $\omega<1/\Delta t$.
will give substantial results. Taking into account that the order of the gravitons numbers value has been
 considerably big we’ll get the condition allowing classical consideration of the averaged field:

\begin{equation}
\vert \vec{G}\vert\gg \frac{\sqrt{\hbar c}}{(c\Delta t)^2}.
\end{equation}

The less is the interval of average $\Delta t$, the more is intensity of the field $\vec{G}$.
For variable fields this interval should not be more than the period of time during which the field varies
 noticeably. Accordingly weak variable fields cannot be quasi-classical, while static field
 is always classical and we can put
 $\Delta t\to\infty$ on for it.

At quantization with the boson sea theory \cite{2},  transition has been performed from gravitons
 of negative energy to gravitons – "holes" – with positive energy. When transiting to the
  classical level it is essential to perform back transition to gravitons with negative energy, particularly,
   to comply with Newton's law of interacting macro bodies. The transition will be discussed
    more detail in the next paragraph. So it seems that two types of gravitational field exist:
     classical one- when masses interaction is defined according to Newton law and quantum one
      – when there is the law of mass repulsion as in electrodynamics for charges of the same sign.

The question is in what distances gravitational field can be considered as quantum and in what distances it is
 classical. It is difficult to answer this question unambiguously. In accordance with the above mentioned
  transition to the classical limit it can be assumed that the field is to be considered as classical in case
   when time derivatives from gravitational field potentials and densities become substantial. The maximum
    distance where gravitational field is considered to be quantum likely exceeds the Planck length by several times.
The transition of laws of microscopic interaction to massive bodies is defined by summing and values averaged
 over particles considerable number for various physical parameters. The main role in forming of macro bodies
  belongs to atom and molecular density, form; atom and molecular interactions caused by electromagnetic forces.
   Summing separate molecules masses give the total mass of the macroscopic body and the mass defines it's
    macroscopic gravitational potential. As it has been already mentioned when transiting from quantum gravitation
     to macro bodies it is essential to consider the change of the gravitational field energy sign
  in addition to summing and average. However, the relevant gravitons - holes gravitino are involved
   only-to the gravitational interaction, and probably have a nonzero but very small mass can be grouped
    together forming gas, characterized by positive-energy, but by the law of the repulsive
   negative pressure. It makes sense to assume that this gas makes main contribution
    in dark energy forming its structural base causing acceleration of Universe.

\section{Vacuum symmetry breaking}

As it has been already mentioned gravitons energy becomes positive at quantization of gravitational field
 in accordance with the boson sea theory. As a result the law of masses interaction changes – it becomes
  equivalent to Coulomb’s law of interaction between charged objects, i.e. the masses begin to repulse.
   Let’s consider the mechanism being a base of this phenomenon. This mechanism is similar to the Higgs
    mechanism but it is not identical to it.

Let us assume that two scalar fields $\varphi_1$ and $\varphi_2$, exist, where the second field can have no mass.
The fields lagrange is the following:

\begin{equation}\label{125}
\L=\frac{1}{2}\partial_\mu\varphi_1\partial^\mu\varphi_1-\frac{m^2}{2}\varphi_1\varphi_1-
\lambda_1(\varphi_1\varphi_1)^2+\frac{1}{2}\partial_\mu\varphi_2\partial^\mu\varphi_2-
\lambda_2(\varphi_2\varphi_2)^2,
\end{equation}

summands containing $\lambda_1$, $\lambda_2$ correspond to self action.
Let’s find the minimum of the potential corresponding to the field $\varphi_1$.
If $m^2>0$, the minimum is at $\varphi_1=0$, if $m^2<0$, the minimum is achieved at

\begin{equation}\label{126}
\varphi_1=\Bigl(-\frac{m^2}{4\lambda_1}\Bigr)^{1/2}=\pm a.
\end{equation}

Each of these solutions can be selected as the main state of the field. Let’s select the main state $-a$.
 The vacuum for $\varphi_2$ is assumed to be nondegenerate.

The summand corresponding to gravitational field energy can be presented as:

\begin{equation}\label{127}
  \L=\frac{1}{8}\frac{\varphi_1}{\varphi_2}
  \frac{\partial A_m}{\partial x^n}\frac{\partial A^m}{\partial x_n}.
\end{equation}

To simplify the theory $\frac{\varphi_1}{\varphi_2}=const$, хalthough in common cases it is not obligatory,
 as this summand is left formally renormalizable. At this interaction between scalar fields and gravitation
  field through the covariant derivative is not expected, so graviton has no mass.

   Considering vacuum medium zero value let’s write $\varphi_1=\varphi_1^{'}-a$ and instead of (\ref{127})
we get:

\begin{equation}\label{128}
  \L=\frac{-a+\varphi_1^{'}}{8\varphi_2}
  \frac{\partial A_m}{\partial x^n}\frac{\partial A^m}{\partial x_n}.
\end{equation}

So due to vacuum degeneracy a summand corresponding to Coulomb gravitation appears. At the quantum level
 within small distances let’s consider $\vert a\vert\gg\vert \varphi_1^{'}\vert$.
The maximum distance where this effect begins to appear can be assumed as exceeding the Planck length by
 several times. In the classical limit vacuum degeneracy cannot be taken into account and gravitation energy
  is (\ref{40}), and accordingly the main gravitational interaction is Newton law.

\section{Dark energy}

According to the modern astronomical observations  \cite{11}, the universe expands with acceleration.
 The cause of this acceleration, as it was established, is the repulsive force of unknown origin,
  called the dark energy.  There is a multiplicity of hypotheses trying to explain the nature of
   dark energy: as the energy of vacuum connected with the cosmological constant  \cite{12},
quintessence \cite{13}, Chaplygin gas  \cite{14}, tachyon field  \cite{15}, phantom scalar fields with negative
 kinetic energy  \cite{16} etc.

As it was shown in  \cite{1} there is a cosmological model of the universe, which
 is equivalent to Friedman model in the framework of vector theory of gravitation.
  In this case, the simple Newtonian picture with elliptic, parabolic and hyperbolic
   motion of the substance depending on the initial speed corresponds to the cosmological
    solutions of stationary, expanding and contracting Universe. Thus, for explanation of
     the astronomical observations of acceleration of the universe expansion, in the framework
      of vector theory of gravitation, proper allowance must be made for dark energy.

  As the dark energy we can consider above-mentioned the Fermi gas of the gravitino with negative pressure.
    If we suppose
    approximately homogeneous distribution of gravitino in the universe, this medium
     coincides with the dark energy by its properties.

The theorem, where the substance surrounding the area under consideration with
 spherically symmetric layer does not affect the processes inside the area one way or the other,
  is true. This statement can be extended to the area in the infinite space,
   filled with the substance with constant density. The theorem is valid both for
    Newton theory, general relativity theory \cite{17} and for vector theory of gravitation.

 Let us consider the spherical area of radius  $a$, inside of which the substance with density $\rho$
has at the moment  $t=t_0$ the speed, distributed under the law:

\begin{equation}\label{90}
 \vec{u}=H\vec{r}.
\end{equation}

Lagrangian of the matter interacting with the gravitational field takes the following form:

\begin{equation}\label{95}
\L=T^{ik}g_{ik}-A_ij^i,
\end{equation}

where

\begin{equation}\label{5x}
T^{ik}=\Bigl(\rho_{U}-\frac{\Lambda}{c^2}\Bigr)\frac{dx^i}{ds}\frac{dx^k}{dt},
\end{equation}

 the tensor of energy-momentum with density of ordinary matter $\rho_{U}$
and dark energy  $\Lambda$. The relation between the tensor of energy-momentum and four-vector momentum
$P^i=\frac{1}{c}\int T^{ik}dS_k$ occurs, integration is made on hypersurface. Four-vector of momentum is
 connected with a four-vector of current of masses  $\int j^idV=P^i$.
As a result, using (\ref{5x}) we get for vector of density of current of masses:

\begin{equation}\label{5y}
j^i=\Bigl(\rho_{U}-\frac{\Lambda}{c^2}\Bigl)\frac{dx^i}{cdt}
\end{equation}

Supplying to  (\ref{95}) the explicit form of tensor of energy-momentum (\ref{5x})
 let us write Lagrange equitations (equitations of motion):

\begin{equation}\label{96}
\frac{du_i}{ds}=G_{ik}u^k,
\end{equation}

where $u^k$ is the four-velocity.
Further, from Poisson equitation (\ref{7}) we find the type of potential  $A_0$ with regard to  (\ref{5y})
for spherical area of radius $a$:

\begin{equation}\label{97}
\varphi_{tot}=-\gamma\int\frac{\rho_{U}}{a}dV_a+\gamma\int\frac{\Lambda}{c^2a}dV_a;
\end{equation}

The result is that when introducing the mass of universe  $M_U=\frac{4\pi}{3}\rho_{U}a^3$
 we write the equitation of motion on radius  $a$ a with neglecting the vector potential:

\begin{equation}\label{98}
\frac{d^2a}{dt^2}=-\gamma\frac{M_U}{a^2}+ \frac{4\pi\gamma}{3c^2}\Lambda a;
\end{equation}

The first term in  (\ref{98}) corresponds to ordinary force of gravity per mass unit, the
 second term corresponds to repulsive force connected with the density of negative energy
  of gravitons. Multiplying both members of the equitation (\ref{98}) by $u_a=da/dt$ we obtain
   ordinary Friedman equitation with cosmological constant $\Lambda$.

\begin{equation}\label{99}
\frac{1}{2}\Bigl(\frac{da}{dt}\Bigr)^2-\frac{4\pi\gamma}{3}\Bigl(\rho-\frac{\Lambda}{c^2}\Bigr) a^2=const.
\end{equation}

 The density of energy of dark matter will be match up the cosmological constant, which is equivalent
  to the density of the Fermi gas of the gravitino with negative pressure.
   Solution (\ref{99}) is specified for the example in  \cite{18}.

\section{Black holes}

For the first time the meaning of a black hole was mentioned by Mitchell and Laplace within Newton theory.
 Total energy of a test body with mass $m$t in gravitational field of a body with mass M is to be defined as the sum of
   kinetic energy and potential energy: $E=\frac{mv(r)^2}{2}-\frac{\Gamma Mm}{r}$.
 If $E\geq 0$ the test body speed shall satisfy the condition

\begin{equation}\label{129}
v(r)\geq \sqrt{\frac{2\Gamma M}{r}}=v_0(r),
\end{equation}

where $v_0(r)$ is the 2nd cosmic velocity. If for a radius $r$ the velocity of
$v_0$ reaches light velocity $c$, no particle including   photon can leave an object with gravitational radius:

\begin{equation}\label{130}
r_g=\frac{2\Gamma M}{c^2}.
\end{equation}

It can be shown that within the vector gravitation theory an effective Schwarzschield metric congruent to black holes can be get.
 Following \cite{19,20} let there is a source of constant symmetrical gravitational field.
  Let’s assume relative space of the infinitely remote observer resting relatively to source is euclidian,
   and so spherical coordinates
 $(r, \theta, \phi)$ with the origin matching with the source center can be introduced.
  Let’s assume that the equivalence principle is fulfilled. Let us suppose that from the point
   of view of the infinitely remote observer the test particle freely falls radially according
    to Newton law of gravitation. The speed of this particle is zero in infinite removal and it is
     equal to the 2nd cosmic velocity in the given radial coordinate:

\begin{equation}\label{131}
v=\sqrt{\frac{2\Gamma M}{r}},
\end{equation}

where $M$ is the source mass.
Let us suppose that $\rho$  is a space coordinate in radial direction of the local inertial reference system
 connected with the abovementioned test particle. This system is connected with the system of the infinitely
  remote observer expressed in terms of the space coordinate $r$ and temporal coordinate $\tau$:

\begin{equation}\label{132}
d\rho=dr-vd\tau,
\end{equation}

The metric of this local system can be written down as follows:

\begin{equation}\label{133}
ds^2=-d\tau^2+d\rho^2+r^2d\Omega,
\end{equation}

where $d\Omega=\sin^2\theta d\phi^2+d\theta^2$.
If we put (\ref{132}) in (\ref{133}) we’ll get:

\begin{equation}\label{134}
ds^2=-\Bigl(1-\frac{v^2}{c^2}\Bigr)d\tau^2-2vd\tau dr+dr^2+r^2d\Omega,
\end{equation}

Next, if we put (\ref{131}) in (\ref{134}) we’ll get:

\begin{equation}\label{135}
ds^2=-\Bigl(1-\frac{2\Gamma M}{rc^2}\Bigr)d\tau^2-2\sqrt{\frac{2\Gamma M}{r}}d\tau dr+dr^2+r^2d\Omega,
\end{equation}

This metric is Schwarzschild metric in the form of Painleve-Gullstrand \cite{21,22,23}.
Let put such temporal coordinate $t$ that:

\begin{equation}\label{136}
dt=d\tau+\vartheta^2vdr
\end{equation}

where $\vartheta=1/\sqrt{1-\frac{v^2}{c^2}}$.  Or equivalent:

\begin{equation}\label{137}
d\tau=dt-\Bigl(1-\frac{2\Gamma M}{rc^2}\Bigr)^{-1}\sqrt{\frac{2\Gamma M}{rc^2}}.
\end{equation}

Putting (\ref{137}) in (\ref{135}) we’ll get effective Schwarzschield metric:

\begin{equation}\label{138}
ds^2=-\Bigl(1-\frac{2\Gamma M}{rc^2}\Bigr)dt^2+\Bigl(1-\frac{2\Gamma M}{rc^2}\Bigr)^{-1}dr^2+r^2d\Omega.
\end{equation}

\section{Binary pulsars}

Observation of the binary pulsar PSR 1913 +16 and the definition of its basic parameters \cite {24} played an important
 role to test the various theories of gravitation.

Double pulsars are complex objects, the nature of many phenomena in which the remains
 not completely unclear. For example in the recent observations of the pulsar PSR J1718-3718
 found unexplained sudden change in the rate of rotation \cite{25} amounting to $33,25*10^{-6}$,
that exceeded the previously observed values of this parameter observed in other pulsars,
and has for more than 700 days the system parameters are not restored.
These changes are associated with enormous energy costs.

Investigation of the binary pulsar PSR 1913 +16 revealed the existence of the effect of reducing energy
 the motion of a pulsar in orbit and as a consequence of reduction of the orbit itself. This effect was associated
 with gravitational radiation. If we assume that the radiation is quadrupole,that
 quantitatively, this effect is consistent with general relativity in a relatively narrow range
  of masses of neutron stars \cite{26}.

Vector theory of gravity is that apart from the possibility of quadrupole radiation
 the dipole radiation.
From the classical equations of the vector theory of gravity should be that the dipole radiation
 have a negative energy, but from show above the quantum theory should that the radiation is still
 must have positive energy.

Without allowance compensating dipole radiation
 in the vector theory of gravitation allowed the existence of the mass of the pulsar and its companion
 satisfying the restriction period changes in the masses of the order $\sim 1.4M_{sun}$
only, or equal value, or nearly equal, differing by only a small amount.
In addition, there can exist a wide range of pulsar masses of $ <\sim 0.4M_ {sun} $ and less
 up to $\sim 0.01M_{sun}$.
The lower boundary of the $m_1$ is defined by corresponds to the change of orbital
 period, assuming a dipole gravitational radiation.

At the same limitations of the Doppler effect and gravitational redshift
 partner mass $m_2$ should be in the range of $> \sim 0.4M_ {sun} $ and about $ <\sim 1.8M_ {sun} $.
In this case, the pulsar may be a so-called pure neutron stars
 pure iron stars or helium stars.

To expand the possible values of the masses in the direction of their increase $>\sim 0.6M_ {sun} $ and the increase
  the width of the mass difference between the partner and the pulsar in frame of the vector theory of gravitation
  we can suggest a mechanism for some compensating dipole radiation,
and thus increasing the energy and orbital period of pulsars.
Such mechanisms can be found a few. The final outcome is likely determined by a combination of these mechanisms.
We list the main ones.
Tidal dissipation \cite{27}. For a companion with rotation axis normal to the planet of the orbit
and companion rotates faster than the orbit by some factor of order unity, dissipation increases
the orbit energy and causes the period to increase. If the source of molecular viscosity is tidally
driven turbulence \cite{28,29}, that for a helium star companion rate of change
 orbital period could then be comparable to the general relativistic quadrupole radiation damping rate.
For a white dwarf companion tidal dissipation is negligible unless the white dwarf is very rapidly
rotation, and a very strong source of viscosity, such as magnetic viscosity is present \cite{30}.

Another partially offset the effect is a third massive body or several bodies.
Also compensating effect is a precession of the pulsar's spin axis \cite{26}.

In addition there is one more important effect, due to the presence of a inhomogeneous cyclic field
 of the binary pulsar. Inhomogeneity may be associated with distance from the center of the pulsar, and so
  the inhomogeneity of the rotation of the pulsar.

By analogy with the force produced in an inhomogeneous magnetic field strength in inhomogeneous cyclic field in a general form
 we can write:

\begin{equation}\label{1202}
\vec{F}=L\frac{\partial \vec {C}}{\partial n},
\end{equation}

where $\vec {C}$ denotes the total induction of cyclic field,
$L$ - momentum, $\vec {n}$ -
 vector in the direction of the normal.
The corresponding change in energy:

\begin{equation}\label{123}
\frac{dE}{dt}=L\Bigl<\frac{\partial \vec {C}}{\partial r}\Bigr > v,
\end{equation}

< > mean an average. The change of the orbital period is as follows:

\begin{equation}\label{124}
\frac{\dot{P}}{P}=-\frac{3}{2E}\frac{dE}{dt}=\frac{3Lav}{Mm}\Bigl<\frac{\partial \vec {C}}{\partial r}\Bigr>,
\end{equation}

where a - semi-major axis, P - orbital period, $M=m_1m_2/(m_1+m_2)$ - the reduced mass,
$m=m_1+m_2$, $E=-\Gamma Mm/2a$. Substituting the values of the parameters of the pulsar PSR 1913+16 \cite{13}
we can obtain an estimate of the inhomogeneity $\Bigl <\frac{\partial C}{\partial r}\Bigr> >\sim 10^{-22} 1/c*m$.
This assessment seems to be quite relevant to reality.

Thus, in the vector theory of gravity can be obtained
  satisfactory explanation of the change in energy of the orbits of binary pulsars.

%%%%%%%%%%%%%%     Литература    %%%%%%%%%%%%%%%%%%%%%%%


\begin{thebibliography}{99}

\bibitem{1}
{\it V.N. Borodikhin.} Grav. Cosmol. 17, 161 (2011).
\bibitem{0}
{\it G.L. Whitrow, and G.E. Morduch.} In Vistas in Astronomy, 6 (Pergamon Press, Oxford, 1965).
\bibitem{00}
{\it C.W. Misner, K.S. Thorne, J.A. Wheeler.} Gravitation (Freeman, San Francisco, 1973).
\bibitem{000}
{\it C.M. Will.} Living Rev. Relativity 4, 4 (2001).
\bibitem{2}
{\it Y. Habara, Y. Nagatani, H.B. Nielsen, M Ninomiya.}  Int. J. Mod. Phys. A23, 2733 (2008);
 hep-th/0603242;
\bibitem{3}
{\it Y. Habara, Y. Nagatani, H.B. Nielsen, M Ninomiya.}  Int. J. Mod. Phys. A23, 2771 (2008);
 hep-th/0607182;
\bibitem{4}
{\it N.N. Bogoliubov, D.V. Shirkov.} Introduction in the quantum fields theory (Nauka, Moscow, 1980).
\bibitem{26}
{\it C.M. Will.} Theory and Experiment in Gravitational Physics (Univ. Press, Cambridge, 1993).
\bibitem{5}
{\it R. Feynman.} Quantum electrodynamics (NFMI, Novokuzneck, 1998).
\bibitem{8}
{\it A.I. Akhiezer, V.B. Berestetskiy.} Quantum electrodynamics  (Nauka, Moscow, 1981).
\bibitem{9}
{\it J.M. Jauch, F. Rohrlich.} The Theory of Photons and Electrons (Springer-Verlag, 1976).
\bibitem{10}
{\it L. Ryder.} Quantum Field Theory (Mir, Moscow, 1987).
\bibitem{11}
{\it S. Perlmutter et al.} Astrophys. J. 517, 565 (1999);
{\it D.N. Spergel et al.} Astrophys. J. Suppl. 148, 175 (2003); astro-ph/0302209;
{\it A.D. Miller et al.} Astrophys. J. Lett. 524, L1 – L4 (1999).
\bibitem{12}
{\it A. Melchiorri} The death of quintessence (In Proc. of the I.A. P. Conf.,
 edit. P. Brax, J. Martin and J.P. Uzan, Frontier Group, Paris, 2002).
\bibitem{13}
{\it C. R. R. Calwell, R. Dave, and Steinhardt.} Phys. Rev. Lett. 80, 1582 (1998).
\bibitem{14}
{\it A. Kamenshichik et al.} Phys. Lett. B 511, 265 (2001).
\bibitem{15}
{\it J. S. Bagla et al.} Phys Rev D 67, 063504 (2003).
\bibitem{16}
{\it R.R. Caldwell.} Phys. Lett. B 545, 23 (2002); astro-ph/9908168;
{\it J. Kujat, R. J. Scherrer, A.A. Sen.} Phys. Rev. D 74, 083501 (2006); astro-ph/0606735.
\bibitem{17}
{\it E. A. Milne and W. H. McCrea.} Q. J. Math. (Oxford) 5, 73, (1934).
\bibitem{18}
{\it J.B. Zel'dovich, I.D. Novikov.} Structure and Evolution of the Universe (Nauka, Moscow, 1975).
\bibitem{19}
{\it  J. Czerniawski.} Concepts Phys. 3, 307 (2006); arXiv:gr-qc/0201037.
\bibitem{20}
{\it  M. Visser.}  Int.J.Mod.Phys. D 14, 2051, (2005); arXiv:gr-qc/0309072.
\bibitem{21}
{\it P. Painlevґe.} C. R. Acad. Sci. 173, 677 (1921).
\bibitem{22}
{\it A. Gullstrand.} Ark. Mat., Astron. Fys. 16, 1 (1922).
\bibitem{23}
{\it W. Israel} in Three Hundred Years of Gravitation, edited
by S. W. Hawking and W. Israel (Cambridge University
Press, Cambridge, England, 1987).
\bibitem{24}
{\it R.A. Huse, J.H. Taylor.} Astrophys. J. 195, 51 (1975).
\bibitem{25}
{\it R.N. Manchester, J. Hobbs.} In press ApJ Letters; astro-ph/1106.5192;
\bibitem{27}
{\it M.E. Alexander.} Astrophys. and Space Sci. 23, 459 (1973).
\bibitem{28}
{\it S.A. Balbus, K. Brecher.} Astrophys. J. 203, 202 (1976).
\bibitem{29}
{\it W.H. Press, P.J. Wiita, L.L. Smarr.} Astrophys. J. 202, 135 (1975).
\bibitem{30}
{\it L.L. Smarr, R. Blandford.} Astrophys. J. 207, 574 (1976).

\end{thebibliography}
\end{document}